# Autonomic Resilient Internet-of-Things(IoT) Management

Rossi Kamal, *Member, IEEE,* Choong Seon Hong, *Senior Member, IEEE,*
and Mi Jung Choi, *Member, IEEE*

## Abstract

In *Resilient IoT*, the revenue of service provider is resilient to uncertain usage-contexts(e.g. emotion, environmental contexts) of Smart-device users. Hence, *Autonomic Resilient IoT Management* problem is decomposed into two subproblems, namely *m-connectivity* and *k-dominance*, such that *m*-alternations on revenue making process is resilient to users common interests, which might be depicted through *k*-1 alternations of usage-contexts. In this context, a greedy approximation scheme *Bee* is proposed, which resolves aforementioned sub-problems with five consecutive models, namely Maverick, Siren, Pigmy, Arkeo and Augeas, respectively. Theoretical analysis justifies the problem as NP-hard, combinatorial optimization problem, which is amenable to greedy approximation. Moreover, *Bee* lays out the theoretical foundation of *Resilient Fact-finding*, followed by theoretical and experimental(i.e synthetic) proof, which show how *Bee*-resilience resolves acute CDS measurement problem. Accordingly, experiments on real Social rumor dataset extract dominator and dominate to justify how Bee resilience improves CDS measurement. Finally, case-study and prototype development are performed on Android and Web platforms in a Resilient IoT scenario, where service provider recommends personalized services for Smart-device users.

## I. INTRODUCTION

### A. Motivation

Internet-of-Things (IoT) is envisioned as the connected world concept, which comprises of Smart-devices, such as Smart-phones, sensors, wearable devices, social network, etc. Consequently, the huge penetration of Smart-devices and increasing demand of personalized services have influenced IoT service providers to earn money from monitoring information. However, extracting common interests/contexts (e.g emotion, geographical information) of users from services is a tedious task. Hence, service providers continuously strive to adapt to usage-dynamics; often seem to motivate user through incentive, social status, qoe-assurance, as conventional tools often seem inadequate in measuring IoT service-usage dynamics.

R.Kmaal is with the Department of Computer Engineering, Kyung Hee University, South Korea. M.J.Choi is with Kangwon National University, South Korea.

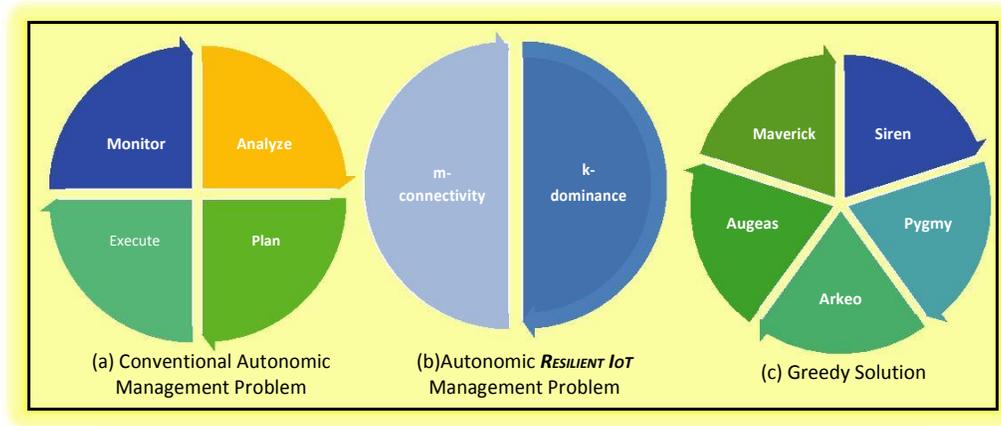

Fig. 1. (a)Traditional Autonomic Management, (b) Autonomic *Resilient IoT Management* Problem (c) Greedy Solution

## B. Relevant Research

Uncertainty of Smart-device traffic has been a major concern among service providers due to deficiencies of poor sensor quality, calibration technique and human-involvement. Especially, uncertainties in quality of submitted information, incompleteness of large corpus of data, recovery of random sample and time dependencies of Smart-device user behavior in different contexts and events are striving service providers for resilient solutions.

An autonomic approach has proliferated to adapt to dynamic environment contexts. IBM's MAPE (Monitor, Analyze, Plan and Execute)-based autonomic approach, is conventionally utilized to Monitor and Analyze Smart device traffic and consequently, Plan, Execute dynamic context-specific network/service management policies. Accordingly, service providers need to adapt his/her autonomous service recommendation policies according to usage-dynamics of Smart-device users. Therefore, the major challenge resides in the way it devises resilient mechanism to resolve uncertain usage-contexts.

## C. Problem Formulation

In this context, Resilient IoT is envisioned, in which the revenue of service providers is resilient to the usage-dynamics of Smart-device users. In this context, service providers seem to deploy various measurement tools (denoted as IF) to extract usage-context (denoted as COI) of Smart-device users. Accordingly, service providers collect real-time/offline feedback of Smart-device users to regulate whether their business goals are met.

Hence, Resilient IoT is aimed at extracting an efficient IF set, which always prevail over preferred COI set. In this context, Resilient IoT is defined as connected dominating set problem, resembled by dominators (e.g IF), who, being connected, dominate over other nodes, denoted as dominates (e.g COI).

However, Autonomic Resilient IoT Management requires service providers to be prepare revenue making process according to usage-context of Smart-device users. In this context, (a) a group of usage-contexts is to be selected, which represents personalized service, (b) a group of revenue making process is to be justified, which extracts usage-contexts of Smart-device users. Hence, Autonomic Resilient IoT Management problem is decomposed into two subproblems, as follows

(a)k-dominance problem: It addresses the extraction of common user-group. Hence, an optimal COI set selection is necessary, who are connected in k-ways to IF for the ease of extraction of common interests.

(b)m-connectivity problem: It addresses the extraction of efficient IF. Hence, an optimal IF set selection is necessary, which are connected in m-ways among each other.

### D. Proposal

In this context, a greedy approximation scheme Bee is proposed, which assists in resilience of service providers revenue to Smart-device users usage-context. Consequently, Bee solves k-dominance and m-connectivity problems with Maverick, Siren, Pigmy, Arko and Augeus models, respectively. In this process, at first MIS is constructed (Maverick) and then nodes are added to it to construct 1-connected 1-dominating set (Siren). It is followed by k-1 MIS addition and iterative augmentation to form 1-connected k-dominating set (Pigmy) and 2-connected k-dominating set (Arkoo), respectively. Finally, bad points are turned to good points to end up with 3-connected k-dominating set (Augeus).

### E. Theoretical Analysis

Theoretical analysis justifies the following-Resilient IoT Service management is an NP-hard and combinatorial optimization problem. However, the absence of polynomial time approximation scheme is justified to necessitate a greedy approximation scheme.

### F. Experiment Results

Extensive numerical analysis on synthetic dataset justifies the following results: as network size or space, packet size, dominator and dominate increase, Resilience IoT Management problem becomes more acute, since CDS size increases. However, the inclusion of resilience assists in gradual improvement in overload, latency and success-ratio. Moreover, experiment on Higgs Twitter dataset yields the following result, (a) 2 pair of latent dominators (b)4 latent dominate are inherent in dataset, which resembles the occurrence of a scientific rumor in the social network. However, as number of dominate or dominate increase, Resilient IoT management becomes complex, however, resilience gradually assists in achieving efficiency in terms of overload, success-ratio and latency.

### G. Organization

The paper is organized as follows-(a)*Autonomic Resilient IoT Management* problem is presented in section 2, (b) Greedy Approximation Scheme *Bee* is presented section 3, (c) Theoretical analysis is presented in Section 4, (c) Empirical results are presented in Section 5, (d) Related work and Conclusion are presented in Section 6 and 7, respectively.

TABLE I

NOTATIONS IN UDG

| Notation | Description |
|----------|-------------|
| G | A graph |
| V(G) | Vertices of a graph |
| (u,v) | A pair of vertices of a graph |
| E(G) | An edge between pair of vertices in a graph G |
| I | Independent Set |
| M | Maximal Independent Set |
| D | Dominating Set |
| UDG | Unit Disk Graph |

TABLE II

UNCERTAINTY FACTORS IN UDG

| Definition | Description |
|------------|-------------|
| Bad point | A vertex v2V(G) is a bad point, if subgraph induced by G-v is not 2-connected. |
| Cut vertex | A vertex veV(G) is a cut vertex if graph G*n* v is disconnected. |
| Leaf block | A leaf block of a connected graph G is a subgraph such that it is a block and contains one cut-vertex of G |
| Independent set | I(G)  V(G) is an independent set of G if *8*(u,v), no edge exists between u and v. |
| Maximal Independent set | An independent set M is a maximal independent set if no v2(G *n* M) can be added to M. If any v2(G-M) is added, it is not an independent set anymore. |

## II. PROBLEM FORMULATION

In this section, *Autonomic Resilient IoT Management* problem is formulated, preceded and followed by preliminary definitions.

Definition 1. *Unit Disk Graph (UDG):A unit disk is a disk having diameter one. A unit disk graph (UDG) is a set of unit disks in Euclidian plane. Each node is in the center of a unit disk. An edge E(u,v) exists between two nodes u and v, if disks associated with u and v intersects each other.*

Definition 2. *D(G) is a dominating set of G if 8v2G, either v2D(G) or 9u such that (u,v)2 E(G).*

Definition 3. *D(G) is a connected dominating set of G if (a) D(G)_V(G) is a dominating set of G and (b) graph*

TABLE III

RESILENCE FACTORS IN UDG

| Definition | Description |
|---|---|
| Good point | A vertex v2V(G) is a good point, if sub-graph induced by G*nv* is still 2-connected. |
| Maximal connected subgraph | A block is a maximal connected subgraph of G that does not have any cut-vertex. |

induced by D(G) is connected.

Definition 4. *A dominating set D(G)_V(G) is k-dominating if 8v2V(G) n D(G), v is adjacent to at least k nodes in D(G).*

Definition 5. *A dominating set D(G)_V(G) is m-connected if graph induced by D is m-connected. It means that D is connected after m* 1 *dominators are removed.*

Definition 6. *Autonomic Management: Given a unit disk graph G(V; E), two positive integers m, k, and uncertainty (Table 2) and resilience (Table 2) components,* Autonomic Management *problem is aimed at finding a subset*

*D _V(G) such that (a)each vertex v 2(V(G)nD) is k-dominated by at least one vertex in D such that guaranteed routing is maintained (b)D is m-connected*

*Hence,* Autonomic Resilient IoT Management *is aimed at facilitating m-connected k-dominating set with mini-mum constraint satisfying (a) and (b).*

## A. ILP formulation

We formulate ILP for *Autonomic Resilient IoT Management* problem. In this context, at first, we formulate ILP for connected dominating set(1)-(11). This portion of ILP formulation is inspired from CDS construction by spanning tree. It is followed by ILP formulation for *m*-connected *k*-dominating set with minimum cost constraint(12)-(15).

$$\min c \tag{1}$$

subject to

$$\sum_{i \in V} a_i \_ D \tag{2}$$

$$b_{ij} \_ na_i; i \in V \; n \; f1g \tag{3}$$

$$\sum_{i \in V} b_{ij} < 1 + (n \; 1) \, a_1 \tag{4}$$

$$\sum_{j \epsilon V} b_{ij} \_ 1 \tag{5}$$

$$\sum_{i,j \epsilon V ; i \neq j} b_{ij} = n \ 1 \tag{6}$$

$$u_i = 1 \tag{7}$$

$$u_j \_ n i \epsilon V \ n \ 1 \tag{8}$$

$$u_j \_ 2 i \epsilon V \ n \ 1 \tag{9}$$

$$u_i \quad u_j + 1 \_ (n \qquad 1)(1 \quad b_{ij}) ; i; j \epsilon V \ n \ 1 \tag{10}$$

$$a_i; b_{ij} \epsilon f0; 1g \ i; j \epsilon V; i \neq j \tag{11}$$

$$\min c \tag{12}$$

subject to

$$\sum_{t \epsilon N(s)} x_t \_ m x_S; \ 8 s \epsilon V \tag{13}$$

$$\sum_{t \epsilon N(s)} x_t \_ k (1 \qquad x_S); \ 8 s \epsilon V \tag{14}$$

$$x_S \epsilon f0; 1g ; \ 8 s \epsilon V \tag{15}$$

Let, $a_i$, $i$ 2V be a binary decision variable indicating whether $i$ beolongs to CDS. Let, $b_{ij}$ , $i,j$2V and i$\neq$j, be a binary decision variable indicating whether that edge is connected to CDS. Let, $c$ represents the cost of CDS.

The objective(1) is to minimize the cost. Constraints are numbered from (2) to (11).(1) means that size of CDS is less than or equal to D.(2) indicates that only the vertices in connected dominating set have outgoing edges.(3)indicates that first node can have edge going out even if it is a leaf node not in CDS. (4)means that first node should have at least one edge going out even if it is a leat.(5)indicates that the spanning tree must have (n-1)edges. (6-11)are used to avoid cycles with inspiration from classical MTZ(Miller,Tucker and Zemlin) formulation.

Let,$x_S$ be a binary variable. $x_S$=1,if s is a dominator. Otherwise, $x_S$=0, if s is a dominate.

 

The objective (12) is to minimize the cost. Constraints are numbered from (13) to (15). The first restriction is, that there exists m disjoint paths between any pair of dominators. Constraint(13) indicates the first restriction. It shows that, if a certain vertex s is set to dominator ($x_S$=1), then vertex s is adjacent to at least m different dominators and thereby form m-connected CDS. The second restriction is that if a vertex is a dominate, then it has at least k dominators. Constraint(14) indicates the second restriction. It shows that if a certain vertex s is a dominate ($x_S$=0), then there exists at least k adjacent dominators to the vertex s. Moreover, constraint(15) indicates that any node s can be either dominator or dominate, so decision variable $x_S$ is either 1 or 0, respectively.

## III. Proposed Scheme

In this section, a greedy approximation scheme *Bee* is proposed, which facilitates the resilience of service providers revenue to uncertain usage-contexts of Smart-device users. In this process, at first diversified revenue-generation sources are extracted, which gradually acquire connectivity and dominance among them and over user-groups, respectively. Hence, Bee resolves *k*-dominance and *m*-connectivity subproblems of *Autonomic Resilient IoT Management* problem with Maverick, Siren, Pygmy, Arkeo and Augeas sub-models respectively as follows. At first (Maverick), isolated revenue-generation sources are identified through MIS construction. Then(Siren), domination of revenue generation source over user-group is determined. Then(Pigmy), diversified domination on user-groups are calculated through 1-connected k-dominating set. However, revenue generation process gradually become connected through iterative augmentation(Arkeo). Finally, dominance of revenue source acquire divergence by converting bad point to good point (Augeas).

## IV. Theoretical Analysis

Theoretical analysis first justifies *Autonomic Resilient IoT Management* problem as NP-hard problem. Eventually, the absence of polynomial time approximation scheme (PTAS) necessitates a greedy solution. Moreover, cost analysis and proof on outcome of different rounds of *Bee* are presented, followed by its differences with existing dominating set research.

Lemma 1. Autonomic Resilient IoT Management is a NP-hard Problem

*Proof.* CDS-construction is a NP-hard problem in UDG[1]. Hence, *m*-connected *k*-dominating set with uncertainty constraints,and thereby *Autonomic Management* is a NP-hard problem. □

Lemma 2. No PTAS exists for *Autonomic Resilient IoT Management* problem

*Proof.* There is no PTAS available for weighted CDS-construction[1]. It yields that no PTAS exists for *m*-connected *k*-dominating set with uncertainty contraint. Hence, no PTAS exists for *Autonomic Management*. □

Lemma 3. *MIS M is created after round 1 of* Bee

*Proof.* In round 1, when a node joins M, its neighbors are colored gray. Next, unexplored white node joins M. So, there is no possibility of gray node to join MIS. So, no two neighbors are included in M. Also, round 1 ends when

TABLE IV

COMPARISON OF *Autonomic Management* PROBLEM WITH PROPOSALS [2],[3] AND [4]

| Pro-posal | Fault-tolerance | Minimum Routing | 3D space Integration | Approxima-tion Bound |
|-----------|-----------------|-----------------|----------------------|----------------------|
| [2] | Present | Absent | Absent | 520/3 |
| [3] | Absent | Absent | Present | 14.937 |
| [4] | Absent | Present | Absent | No PTAS exists. |
| AVIDO | Present | Present | Absent | No PTAS exists. |

there is no white node. So, there cannot be any node left to be added to MIS M. So, MIS M is created after round 1 of *Bee.*

$\square$

**Lemma 4.** *CDS D is created after round 2 of Bee.*

*Proof.* At round 2, all intermediate grey nodes in the path of two MIS nodes(where ROUTINGCOST is below THRESHOLD) are colored black. Those black nodes (intermediate nodes and MIS nodes) construct connected dominating set. $\square$

**Lemma 5.** *1-connected k-dominating set is created after round 3 of* Bee

*Proof.* Let, G, D, I be connected graph, connected dominating set and maximal independent set respectively. After MIS and then CDS construction in first two rounds, k-1 subsequent MIS are added to CDS in third round. As a result of it, G-D nodes are k dominated by D nodes. That means each node of G-D is connected to k nodes of D. So, 1-connected k-dominating set is found. $\square$

**Lemma 6.** *2-connected dominating is created after round 4 of* Bee. Proof: *At the end of round 4, all dominator nodes are in same block, so that dominators are 2-connected[5]. So, we get 2-connected k-dominating set at the end of round 4.*

**Lemma 7.** *If v$\epsilon$G is a good point, subgraph G- fvg is 2-connected[5].*

**Lemma 8.** *A 2-connected graph without any bad point is 3-connected.*

*Proof.* A graph G is 3-connected if we need to remove at least three nodes to disconnect G. For example, v be a good point in 2-connected graph G'. From Lemma 11, G'-(v) is still 2-connected. So, we need to remove at least 2 nodes to disconnect G'. So, we can say G' is 3-connected. $\square$

**Lemma 9.** *3-connected dominating is created after round 5 of* Bee

*Proof.* In round 5, all bad points are converted to only good points by transferring some nodes from non-dominator set to dominator set. At the end, 3-connected 3-dominating set is created [2]. □

**Lemma 10.** *Approximation bound for uncertainty constraint in* Bee *is* $d(u; v) \_ 4$.

*Proof.* We have considered uncertainty as a constraint in *Bee*.

Let, $u$, $v$ are two nodes to be connected to CDS with minimum uncertainty and $d(u; v)$ is distance between them. Threfore, $d(u; v) \_ 4$ is obtained by following[4] and by considering the following facts.

Let, $a; b$ are two MIS nodes such that $d(a; b) = 2$, it means that no MIS nodes are more than two hops from each other. Now, $u; v$ are two nodes such that $d(u; v) \_ 4$. Let us connect $u; v$ to connected dominating set. Therefore, $u$ can be one hop away from $a$ and $v$ can be one hop away from $b$.

Therefore, $d_D(u; v) <= d(a; b) +$

$2 \ d_D(u; v) \_ 2 + 2 \_ 4$
$d_D$ represents the distance between $u; v$ when they are connected in *CDS*.

□

## V. EMPIRICAL RESULTS

### A. Numerical Analysis

The performance of the proposed mechanism is measured by extensive numerical analysis in a Sinalgo simulator. At first, CDS size is measured for different network size, transmission range and uncertainty cost, etc in a random topology. Then, resilience is measured through different network size, density, packet size and even variable dominator, dominate for three parameters (i.e overhead, latency and success-ratio) in both grid and random topologies.

*1) Simulation Settings:* Sinalgo simulator is chosen, as it supports unit disk graph network models and is also regarded as prominent tool in graph theory in devising connected dominating set. The performance of the proposed mechanism is evaluated in terms of major performance metrics (i.e. CDS size, maximum routing length, network size, packet chunk, density and network topology and variance of dominator or dominate). Hence, an empirical packet loss model is considered in our simulation, where packet chunk depends on distance and among nodes.

*2) CDS Size Measurement:* In this experiment, nodes are randomly deployed in a 100X100 plane. The number of nodes range from 50 to 150. One hundred connected UDGs are randomly generated in this simulation setup. All nodes are assumed to have same transmission range. A random value between 0:0 and 0.8 is assigned as the transmission rate between the nodes.

Fig. 1(a) shows how CDS size changes with the transmission range and network size. As the transmission range increases, the CDS size decreases because CDS nodes dominate more non-CDS nodes and fewer nodes are needed to construct the CDS. As the network size increases, the CDS size increases as a larger CDS is needed to dominate the non-CDS nodes.

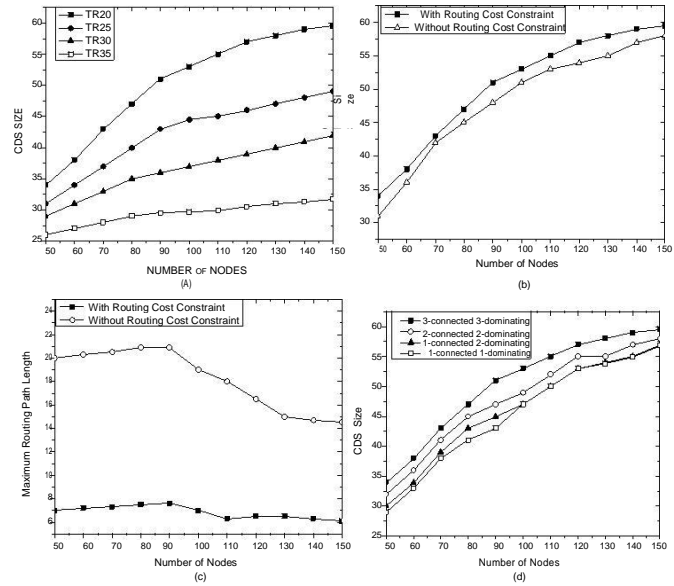

Fig. 2. CDS Measurement by varying(a)nodes for different transmission ranges, or (b)uncertainty cost constraint, by considring presence of (c) maximum cost or (d) resilience

Fig.1(b) shows the impact of uncertainty constraint on CDS size. When network size increases, With uncertainty cost constraint generates larger backbone than without uncertainty cost constraint. Because, it needs more nodes to add to CDS to generate shortest path in CDS for node pairs outside CDS.

Fig. 1(c) shows the impact of uncertainty constraints on the maximum cost. With uncertainty cost generates less maximum cost than without uncertainty cost. When using with uncertainty cost, a node has high probability of connecting to more neighbors,which does not increase uncertainty cost. Therefore,uncertainty cost constraint increases the backbone size, but decreases the maximum uncertainty cost for every node.

Fig. 1(d) shows how CDS size is changed with change in resilience. When, CDS has no resilience(1-connected 1-dominating set), CDS size is the minimum. Gradually, when to make 1-connected 2-dominating set, we need to add one more MIS to CDS. So, with improvement of resilience, CDS size is increased as well. When we make 2-connected 2-dominating set, we need to augment the backbone by adding nodes to connect leaf block in the backbone to otherblock/blocks. As a result, CDS size increases. When we make 3-connected 3-dominating set, CDS size increases as well. There are two reasons for that. Firstly, non-CDS nodes have to move to CDS to convert bad points to good points. Secondly, more MIS nodes have to be added to CDS.

*3) Resilience Measurement: Simulation Settings:* In this experiment, resilience of the proposed mechanism is measured in both grid and random network topologies, where nodes are assumed to be distributed in a rectangular grid and uniformly, respectively. However, the grid network is denoted by m*n-s, where m and n are dimension of rectangular grid and s is the distance between node and closest neighbor. Accordingly, resilience is calculated for three major simulation parameters (i.e. overload, latency and success ratio) by varying network size, packet size and density in a grid topology. The overhead of any resilience is measured in terms of total number of packets involved in CDS construction. However, latency is the time from the starting point of core CDS construction ($m$-connectivity) until the last packet transfer from dominator to dominate ($k$ dominancy). Moreover, success ratio is the ratio of successful packet transfer, which facilitate $m$ connectivity and $k$-dominancy among nodes.Consequently, relative performance is also measured in the random network topology in terms of aforementioned three metrics.

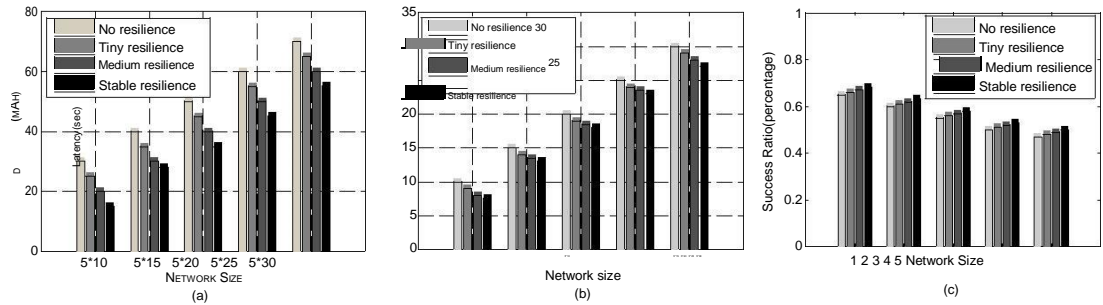

Fig. 3. Resilience measurement by varying network size through (a)communication overhead, (b)latency, (c) success ratio in a grid network topology

*Impact of Network Size:*Fig. 2 shows the impact of network size on resilience measurement. As the network size increases (i.e. n is varied from 10 to 30), CDS size increases from 9 to 22. Thus, dominators remain 15-30 per cent of total nodes on average. As network size increases, both overhead (Fig. 2(a)) and latency (Fig. 2(b)) increases, however, success-ratio (Fig. 2(c)) decreases. Moreover, latency increases slightly slower than overhead with network size, since after the creation of CDS, the packet delivery from dominator to dominate remains almost same. However, in both cases, the gradual inclusion of resilience results in 5 to 10 per cent decrease and increase in overhead-latency and success-ratio respectively.

*Impact of Packet Size:*Fig.3 shows the impact of packet size on resilience measurement. As the packet size increases, packet transfer from dominator to dominate incurs larger overhead (Fig. 3(a)), delay (Fig. 3(b)) and yet decreased success-ratio (Fig. 3(c)) gradually. Because, after CDS is constructed, there is an increase in the number of packets dominators need to send to dominate. However, communication overhead and latency gradually decrease as resilience increases, which becomes substantial for larger packets. Moreover, as resilience increases success ratio gradually increases, since overhead for dominator selection is reduced gradually during data dissemination period.

*Impact of Network Density:*Fig. 4 shows the impact of network density in resilience measurement. As network density (i.e. space between nodes) decreases, overhead (Fig. 4(a)) and latency (Fig.4(b)) are increased and success-ratio (Fig. 4(c)) is decreased gradually. Because, nodes have few good neighbor nodes, which incur visiting more nodes to construct dominating set. It is observed that, as the spacing between nodes increase from 5 to 25 cm, the dominator nodes increase from 15 to 35 per cent. Moreover, when the spaces are higher enough, there is also substantial increase in average packet loss. However, in all scenarios (i.e both communication overhead, latency and success-ratio),10-20 per cent improvement is achieved, as resilience gradually increases.

*Impact of Random Network Topology:*Fig. 5 shows the impact of random network topology in resilience measurement. In aforementioned cases, protocol performances are compared in a grid network topology. Accordingly, its relative performance in a random network topology are measured, where 100 nodes are uniformly distributed in 50*100 area. It is observed that obtained results are almost similar to grid network topology in cases of communication overhead (Fig. 5(a)), latency (Fig. 5(b)) and success-ratio (Fig. 5(c)).

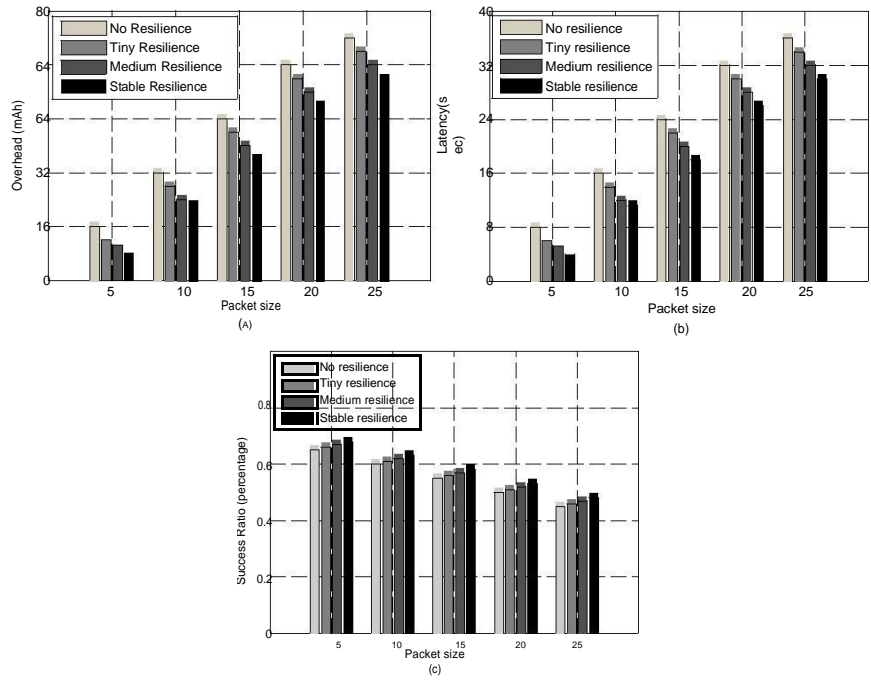

Fig. 4. Resilience measurement by varying packet size through (a)communication overhead, (b) latency, (c) success ratio in a grid network topology

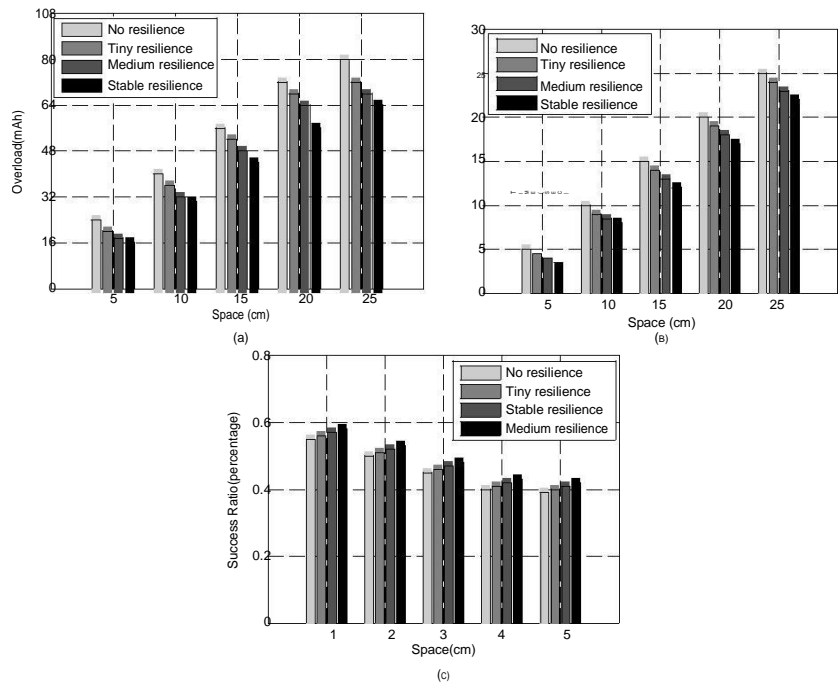

Fig. 5. Resilience measurement by varying network density through (a)communication overhead, (b) latency, (c) success ratio in a grid network topology

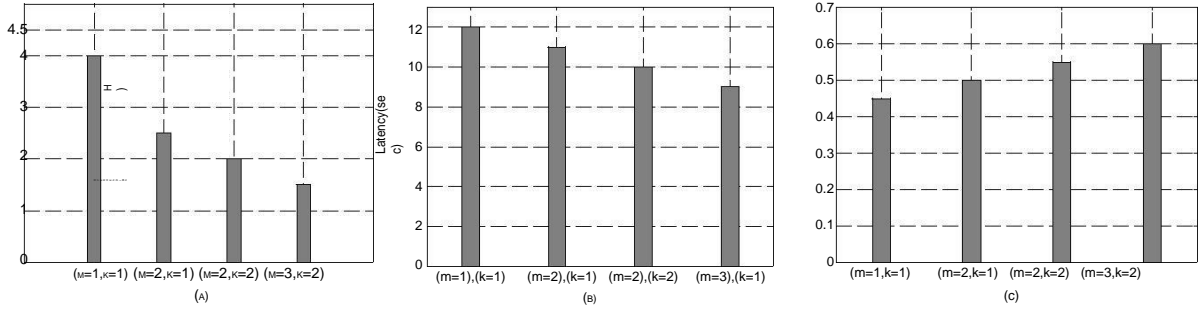

Fig. 6. Resilience measurement by varying network density through (a)communication overhead, (b) latency, (c) success ratio in a grid network topology

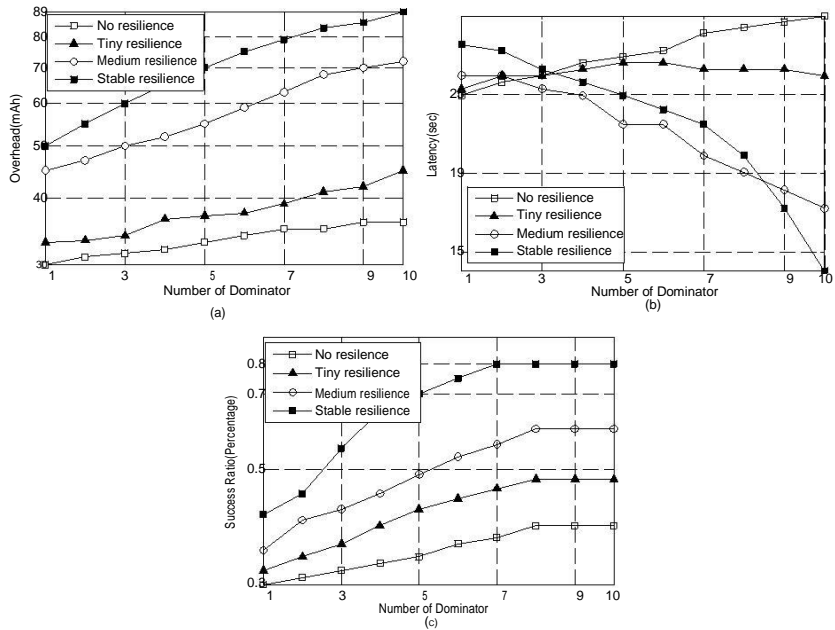

Fig. 7. Resilience measurement by varying number of dominator through (a)communication overhead, (b) latency, (c) success ratio in a grid network topology

*Impact of Dominator:*Fig. 6 shows the impact of dominator on resilience measurement, when dominate are kept constant. As dominator increases, communication overhead increases slightly, however latency and success-ratio are improved gradually. Because, the inclusion of more dominator makes CDS construction easier, which leads to but improved latency and reduced packet failure-rate.However, as resilience increases, more dominators avail themselves to respond to construct CDS with dominate and therefore, event overhead increases. Meanwhile, latency and success ratio are improved, since the average distance between dominator and dominate decreases.

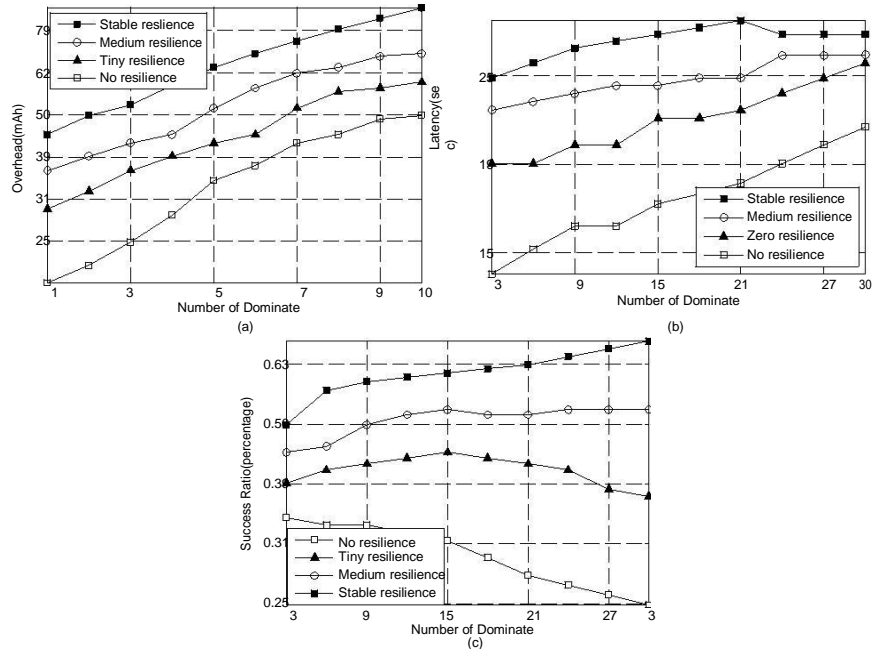

Fig. 8. Resilience measurement by varying number of dominate through (a)communication overhead, (b) latency, (c) success ratio in a grid network topology

*Impact of Dominate:*Fig. 7 shows the impact of dominate on resilience measurement, when dominator are kept constant. As dominate increases, gradual degraded performance is achieved with communication overhead, latency and success-ratio, respectively. Because, the more dominate appear in the scenario, the more is communication load or latency or chance of packet-failure in fixed number of dominators for CDS-construction. However, as resilience increases, gradually better performance is achieved with all cases. Because, the average communication between dominators and dominate become more closer, especially when multiple dominates use same dominator or the vice-versa.

## VI. CONCLUSION

In *Resilient IoT*, the revenue of service provider is adaptive to uncertain usage-context (e.g. emotion, geographical information) of Smart-device users. Hence, *Autonomic Resilient IoT Service Management* is formulated as *m*-connected *k*-dominating set, resembled by dominator (i.e service provider) and dominate (i.e smart-device user), such that service providers are m-connected (i.e who are connected until m-1 alternative ways) to k-dominating users (i.e who are connected n-1 alternative ways). Consequently, a greedy approximation scheme *Bee* is proposed, which facilitates m-connected k-dominating set with an uncertainty constraint. Theoretical analysis justifies the problem as uncertainty, NP-hard and combinatorial optimization problem, alongside with its amenability to greedy asymptotic solution. Extensive numerical analysis on synthetic dataset justifies CDS and resilience measurement efficiency of the proposed scheme in terms of network size, density, packet size, dominator and dominate. Moreover, experiments on three real dataset (i.e Higgs Twitter, CenceMe and YouTube-usage) extract uncertain dominator, dominate and therefore measure CDS and resilience in terms of overhead, success-ratio and time. Finally, case-study and prototype-development are performed on Android and Web-platform in Resilient IoT scenario, where Smart service recommendation (e.g. browsing, instant messaging, social-networking, etc.) is resilient to uncertain usage-contexts (e.g different part of a day, emotion, weather and location).

---

**Algorithm 1 Bee**

---

1. Round 1: MIS Construction

2. INPUT: Color all nodes as WHITE node

3. Choose a node with maximum cardinality and select as root of MIS and color it as BLACK.

4. Color the neighbors of MIS node as GREY node

5. while There is no WHITE node do

6.      Choose the WHITE node that has the most grey neighbors and color it BLACK as MIS node

7.      Color the neighbors of new created black node as GREY

8. end while

9. OUTPUT: BLACK MIS nodes and other GREY nodes

10. Round 2: 1-connected 1-dominating set construction

11. INPUT: initially D is empty and BLACK MIS nodes, GREY nodes are present

12. for Every pair of BLACK nodes u and v with d(u,v)<=4 do

13.      Compute shortest path p(u,v) and color all intermediate GREY nodes of p(u,v) as BLACK

14.      Add u, v and intermediate nodes to D

15. end for

16. OUTPUT: D contains all BLACK nodes (MIS and connected nodes)

17. Round 3:1-connected k-dominating set

18. INPUT: 1-connected 1-dominating set

19. Remove MIS from the graph,G= G-$M_1$

20. for i=2 to k  do

21.      Construct $M_i$ in G-$M_1$ $[$ $M_2$ $[$ ...$M_i$   1

22.      D= D $[$ $M_i$ (Following Round 2)

23. end for

24. OUTPUT:1-connected k-dominating set

25. Round 4:2-connected k-dominating set

26. INPUT:1-connected k-dominating set

27. Find all blocks in 1-connected k-dominating set

28. while D is not 2-connected do

29.      Compute all blocks in graph

30.      Add all intermediate nodes of shortest path that (a)connects leaf block in D to other part of D
             (b)does not have any nodes in D except two endpoints

31. end while

32. OUTPUT:2-connected k-dominating set

33. Round 5: 3-connected k-dominating set

34. INPUT:2-connected k-dominating set

35. while There is no badpoint do

36.      Convert bad point to good point by moving from G-D to D, such that no new bad point is created

37. end while

38. OUTPUT:3-connected k-dominating set

---